\documentclass[aps,prl,twocolumn,superscriptaddress,nofootinbib]{revtex4-1}
\usepackage{amsmath}
\usepackage{graphicx}
\usepackage{amsbsy}
\usepackage{amssymb}
\usepackage[none]{hyphenat}
\usepackage{tikz}
\usepackage{xy}
\usepackage[none]{changes}
\usepackage{wrapfig}
\usepackage{float}
\usepackage{mciteplus}
\usepackage{tabularx}
\usepackage{natbib}
\usepackage{xcolor}

\begin{document}

\title{Leveraging viscous peeling in soft actuators and reconfigurable microchannel networks}

\author{Lior Salem}
\affiliation{Technion Autonomous Systems Program, Technion $-$ Israel Institute of Technology, Haifa, Israel}
\author{Benny Gamus}
\affiliation{Faculty of Mechanical Engineering, Technion $-$ Israel Institute of Technology, Haifa, Israel}
\author{Yizhar Or}
\affiliation{Technion Autonomous Systems Program, Technion $-$ Israel Institute of Technology, Haifa, Israel}
\affiliation{Faculty of Mechanical Engineering, Technion $-$ Israel Institute of Technology, Haifa, Israel}
\author{Amir D. Gat}
\email[Corresponding author: ]{amirgat@technion.ac.il}
\affiliation{Technion Autonomous Systems Program, Technion $-$ Israel Institute of Technology, Haifa, Israel}
\affiliation{Faculty of Mechanical Engineering, Technion $-$ Israel Institute of Technology, Haifa, Israel}

\date{\today}

\begin{abstract}
The research fields of microfluidics and soft robotics both involve  complex small-scale internal channel networks, embedded within a solid structure. This work examines leveraging viscous peeling as a mechanism to create and activate soft actuators and microchannel networks, including complex elements such as valves, without the need for fabrication of structures with micron-scale internal cavities. We consider configurations composed of an internal slender structure embedded within another elastic solid. Pressurized viscous fluid is introduced into the interface between the two solids, thus peeling the two elastic structures and creating internal cavities. Since the gap between the solids is determined by the externally applied pressure, the characteristic size of the fluidic network may vary in time and be much smaller than the resolution of the fabrication method.  This work presents a model for the highly nonlinear elastic-viscous dynamics governing the flow and deformation of such configurations. 
Fabrication and experimental demonstrations of micron-scale valves and channel-networks created from millimeter scale structures are presented, as well as the transient dynamics of viscous peeling based soft actuators. The experimental data is compared with the suggested model, showing very good agreement. 
\end{abstract}
\maketitle

The term \textit{viscous peeling} denotes fluid propagation into the interface between two deformable solids which are initially in contact with each other. In this work, we analytically and experimentally study viscous peeling as an approach to create and activate soft actuators and reconfigurable microchannel networks. 

\begin{figure*}[t] 
\centering
\includegraphics[scale=1]{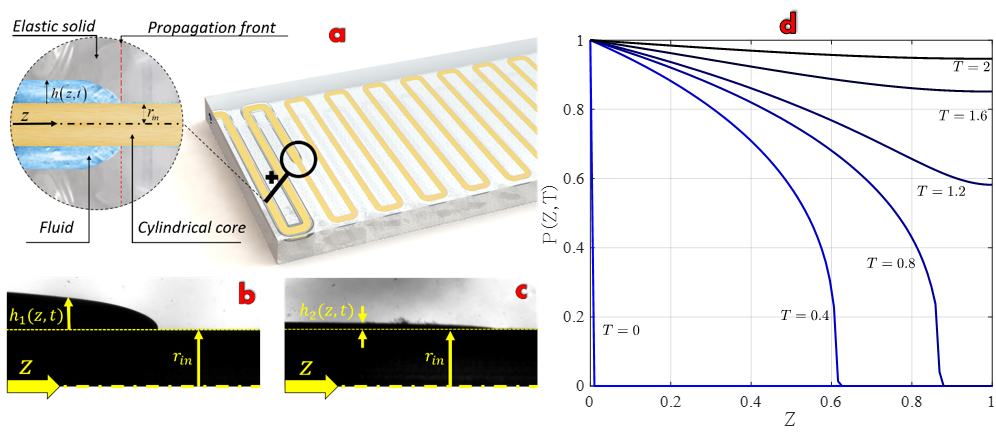}
\caption{Viscous peeling of an embedded slender cylinder. \textbf{(a)} Illustration of the deformation and flow propagation front due to fluidic pressurization of the inlet.  \textbf{(b)} Experimental figure of the propagation front shape without initial pre-wetting layer, and \textbf{(c)} with initial pre-wetting layer. \textbf{(d)} Numeric solution of equation \ref{eq:final} showing pressure propagation along the cylinder for various times. This solution presents non-smoothness at a distinct propagating front, characterizing viscous peeling type dynamics. This non-smoothness is presented for $T=0.4$ and $T=0.8$ and create a sharp separation between pressurized and unpressurized parts of the channel.}
        \label{fig:ModelFull}
\end{figure*}

Fluidic problems involving viscous peeling type dynamics are encountered in the context of biological flows \cite{grotberg2004biofluid}, geophysical and geological phenomena,  \cite{michaut2011dynamics,bunger2011modeling,hewitt2015elastic,thorey2016elastic} as well as oil and gas recovery  \cite{makogon2007natural,lai2016elastic}. The dynamics of viscous peeling are commonly governed by parabolic PDEs \cite{young2017long,hosoi2004peeling,thorey2016elastic,elbaz2016axial}, such as the Porous-Medium-Equation \cite{vazquez2007porous}. These equations involve an inherent nonlinearity, which often yields compactly supported solutions with a non-smoothness at a distinct propagating front (see Figure \ref{fig:ModelFull}(b)), similarly to the dynamics of free-surface flows \cite{oron1997long} and gravity currents \cite{huppert1982propagation}.

This work focuses on leveraging viscous peeling to simplify the fabrication and increase the versatility of embedded channel networks, which is relevant to lab-on-a-chip devices as well as to  the emerging field of soft-robotics \cite{ilievski2011soft,onal2016system,onal2017soft}. By forced introduction of fluid into the interface between a slender inner solid and an external surrounding solid, viscous peeling can be used to create solid structures containing internal networks of fluid-filled channels and chambers. This eliminates the need to remove the inner core within an elastic structure to achieve internal cavities \cite{Marchese2015,Saggiomo2015,Esser2011}. In addition, the nonlinear compactly supported deformation patterns evolving in such configurations allow the creation of sharp separation between activated parts and non-activated parts of the soft actuator. Thus, multiple deformation modes can be achieved by a single geometrical configuration.

Another property of such configurations is that the internal gap between the solids, created by pressure-driven flow, is proportional to the applied fluidic pressure. Thus, for sufficiently small pressures, the thickness of the channel network created in these configurations can be much smaller compared with the characteristic length scales of the solids before the introduction of fluid. This allows to create micron-scale fluidic channels (e.g. for lab-on-a-chip devices \cite{leslie2009frequency,mosadegh2010integrated}) from millimeter-scale structures, and may pave the way to production of sub-micron scale fluidic channels based on the same principle. In addition, this approach readily allows for the fabrication of complex geometries, which is required for micro-fluidic lab-on-a-chip components such as onboard valves.

The following sections model and experimentally demonstrate the incompressible viscous peeling type flows in soft actuators and microfluidic devices. The derivation of  governing equations for propagation of viscous peeling is presented, as well as solutions for simplified cases.  We then present experimental results of various micro-channel configurations and valves, as well as the transient response of a viscous peeling based soft actuator to fluid pressurization. These experimental measurements are compared with the theoretical model.

\section*{A model of the viscous peeling process}
In order to understand the expected evolution of the viscous flow-field and elastic deformation, this section presents the derivation and solutions of the nonlinear diffusion equation governing the process of viscous peeling. 

For concreteness, we consider a slender cylinder of radius $r_{in}$ and length $l$ contained within a second elastic material (as illustrated in Figure \ref{fig:ModelFull}(a)), though other geometries can be similarly modeled. A fluid with viscosity $\mu$ and density $\rho$ is introduced into the interface between the rigid inner cylinder and the outer elastic material. The cylindrical coordinates used are $(r,z)$, fluid velocity is $\mathbf{u}=(v,w)$ in respective directions, and fluid pressure is $p$. Body forces are neglected and the flow is assumed to be axi-symmetric. 

The incompressible Newtonian fluid is governed by the Navier-Stokes equations $\rho \text{D} \mathbf{u}/{\text{D} t}=\mu \nabla^2 \mathbf{u}-\nabla p
$ and conservation of mass $\nabla\cdot\mathbf{u}=0$. Applying the lubrication assumptions \cite{Leal2007} of negligible inertial effects and slender geometry ($v \ll w$, $h^* \ll l$, $h^*$ is characteristic gap) yields the scaling of time $T=t/(\mu l^2/E{r_{in}^2})$, velocity $(V,W)=(v/(E{h^*}^2/l\mu),w/(Eh^*/\mu))$, pressure  $P=p/E$, and coordinates $Z=z/l$, and $R=r/r_{in}=1+\sigma_r\eta$ (for $\sigma_r=h^*/r_{in}$), where $\eta(Z,T)=h/h^*$ is the normalized thickness of the fluidic layer.

Integral mass-conservation equation in these normalized coordinates is
\begin{equation}\label{eq:imass}
    2\pi \left( 1+{{\sigma }_{r}}\eta  \right)\frac{\partial \eta }{\partial T}+\frac{\partial Q}{\partial Z}=0,
\end{equation}
where $Q=q/(Eh^{*3}/\sigma_r\mu)$ is the normalized volumetric flux. Applying the standard lubrication approximation on the Navier-Stokes equations gives
\begin{equation}\label{eq:Q}
  Q \left( Z,T \right)  = 2\pi \int\limits_{0}^{\eta}{W \left( 1+\sigma_r\tilde{R} \right) d\tilde{R}} 
 =\frac{\pi}{8{\sigma_r^3}}\frac{\partial P}{\partial Z} F(\eta).
\end{equation}
where
\begin{equation}\label{eq:Feta}
        F\left( \eta  \right) \triangleq \frac{ \left[\left( \eta\sigma_r+1  \right)^2-1\right]^2}{\ln \left( 1+\sigma_r \eta  \right)}-\left( \eta\sigma_r+1  \right)^4+1.
\end{equation}
An additional relation between the pressure $P$ and the elastic radial deformation $\eta$ is required to fully formulate the problem. Experimental results suggest a linear relation between the pressure $ P$ and the change in channel area (see data in SI and Figu.\ref{fig:dadp}). This linear relation is expressed by   
\begin{equation}\label{eq:constitutive}
    A P=\left( 1+\sigma_r\eta \right)^2-\left( 1+\sigma_r\eta_0 \right)^2,
\end{equation}
where $A=a/(\pi r_i^2/E)$ ($a$ is the dimensional constant) is a dimensionless constant of proportionality determined for each configuration and $\eta_0$ is the initial gap at $P=0$. Combining equations \ref{eq:imass}-\ref{eq:constitutive} yields 
\begin{multline}\label{eq:final}
    8A\frac{\partial P}{\partial T}=\frac{\partial }{\partial Z}\Bigg( \Bigg\{ \left[A P+(1+\sigma_r\eta_0)^2\right]^2-1\\-2\frac{ \left[A P+(1+\sigma_r\eta_0)^2-1\right]^2}{\ln \left[ A P+(1+\sigma_r\eta_0)^2\right]} \Bigg\}\frac{\partial P}{\partial Z} \Bigg),
\end{multline}
which is a strongly nonlinear PDE of the normalized pressure $P(Z,T)$. Equation \ref{eq:final} is supplemented by the inlet pressure boundary condition  $ P(0,T)=P_{in}(T)$ as well as sealed boundary at $Z=1$,   ${\partial P}/{\partial Z}(1,T)=0$, and an initial condition  $ P(Z,0)=0$. For the general case, equation \ref{eq:final} is only solvable numerically. For $\eta_0=0$, the zero diffusion coefficient yields propagation with a compactly-supported non-smooth front, as evident in the numerical solution presented in Figure \ref{fig:ModelFull}(d). The numeric solution was obtained via the MATLAB numeric solver PDEPE, with linear interpolation at the front point.  For the limit of small characteristic gap relative to cylinder radius, $\sigma_r \ll 1$, equation \ref{eq:final} is simplified to
\begin{equation}\label{PME4}
    \frac{\partial P}{\partial T}= \frac{A^2}{48} \frac{\partial}{\partial Z}\left( P^3\frac{\partial P}{\partial Z} \right).
\end{equation}
which is a nonlinear diffusion equation of the form of a fourth-order Porous Medium Equation. Such equations allow for self-similarity solutions \cite{vazquez2007porous,Leal2007} for a sudden release of mass at the inlet, 
\begin{equation}\label{transient}
    P\left( Z,T \right)=T^{-1/5} \left( C-\frac{3}{40}Z^2 T^{-2/5} \right)_+^{1/3}
\end{equation}
where $(f)_+=max(f,0)$ and thus the front location is obtained by setting $P=0$. Ahead of the front, trivial solution of uniform zero gauge pressure is obtained.  For sudden release of mass $M$ at the inlet given by $P(Z,0)=2M\delta(Z)$, where $\delta$ is Dirac's delta function, $C$  is obtained from $\int_0^\infty{P\text{dZ}}=M/A$.

In the following sections, steady state solutions of equation \ref{eq:final} are compared with experimental data of flows in various viscous peeling based microchannel networks. Transient solutions \ref{transient} are compared with experimental data of  viscous peeling based soft actuators.

\section*{Viscous peeling in microfluidic networks and valves}
Since the gap between the solids is proportional to the fluid pressure, the characteristic thickness of the formed cavities can be much smaller compared with the fabrication resolution. In addition, the inner core geometry may be complex, and the creation process of the fluid-filled cavities would not require removing gas bubbles. Thus, such viscous peeling dynamics may be of interest to the fields of microfluidics and lab-on-a-chip devices. This section experimentally studies viscous peeling based micron-sized channels and valves relevant to microfluidics, which are created from millimeter-scale structures.

\begin{figure*}
    \centering
    \includegraphics[scale=0.75]{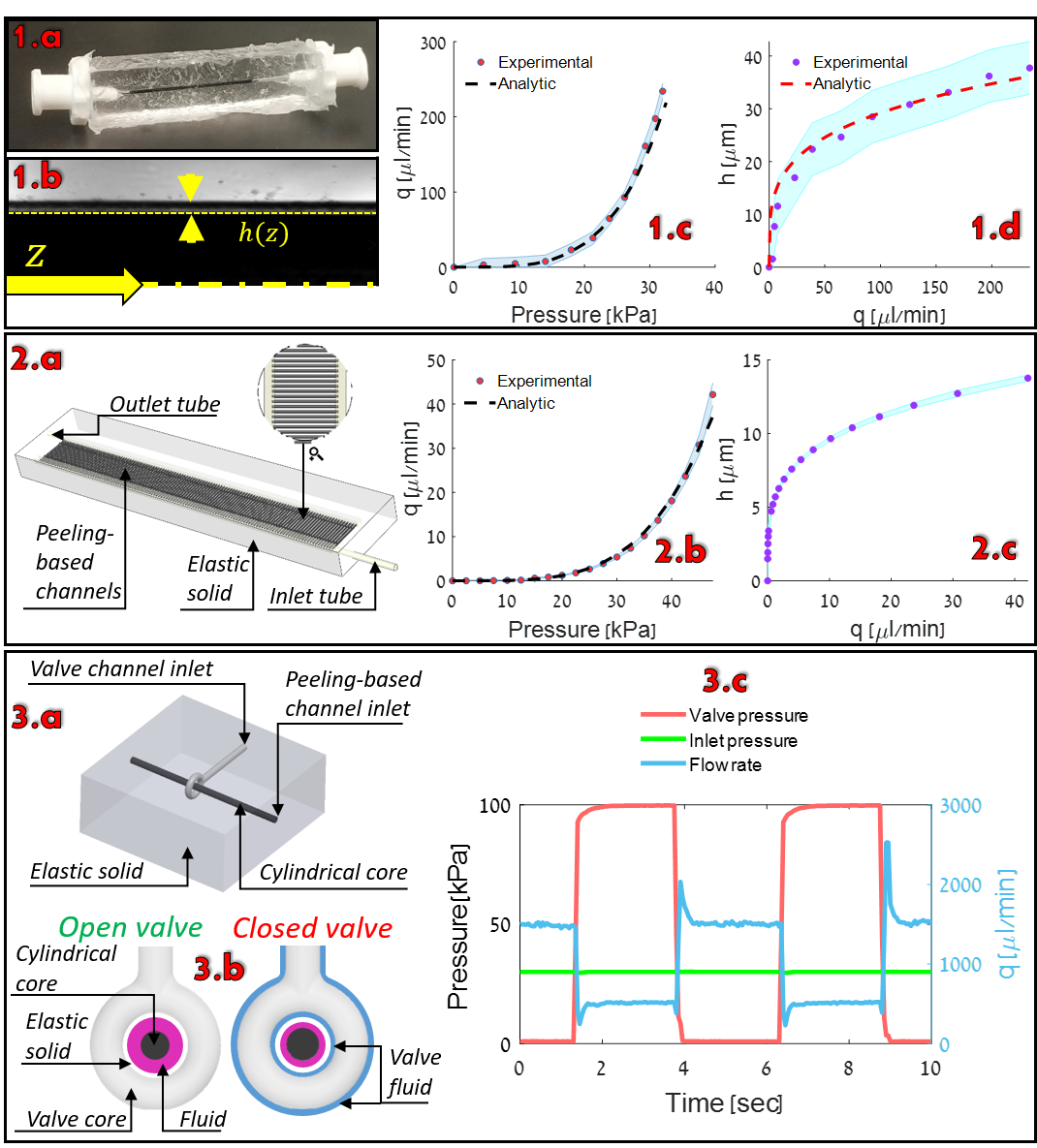}
    \caption{Viscous peeling in microfluidic networks and valves. \textbf{(1)} Experimental illustration of leveraging viscous peeling to create a single microfluidic channel. \textbf{(1.a)} Single channel microfluidic device. \textbf{(1.b)} Gap thickness photo used to measure the gap thickness at the center of the channel. \textbf{(1.c)} Inlet pressure vs. volumetric flow rate measurements. Red dots denotes mean measured values and black dashed lines are analytic calculations. Shaded blue region represents $1$ standard deviation.  \textbf{(1.d)} Gap thickness at $z=L/2$ vs Volumetric flow rate. Purple dots denotes mean measured values and red dashed lines denote analytic calculations. Shaded cyan region represents $1$ standard deviation. \textbf{(2)} Experimental illustration of peeling-based channel network. \textbf{(2.a)} Illustration of the peeling-based channel network configuration. \textbf{(2.b)} Experimental measurements (dots) and analytic calculations (dashed line) of flow rate vs inlet pressure. \textbf{(2.c)} Gap thickness at $z=L/2$ versus flow rate estimated analytically from experimental values of flow rate. \textbf{(3)} Embedded peeling-based valve for microfluidic devices. \textbf{(3.a)} Illustration of the device consisting of a main peeling based channel, and a ring-shaped viscous peeling valve, which are both embedded in elastic structure. \textbf{(3.b)} Cross section of peeling-based channel and ring valve. Left figure illustrates "Open valve" state, were the valve channel is unpressurized. Consequently, fluid flows without restriction along the main channel. Right figure illustrates partially "Closed valve" state, were the valve channel is pressurized. Valve's fluid force the elastic material to restrict the flow in the main channel. \textbf{(3.c)} Experimental results of flow rate through peeling-based main channel vs valve inlet pressure. Flow rate of the peeling based main channel drops as we pressurized the valve channel.}
    \label{fig:Microfluidics}
\end{figure*}

Three different configurations were examined, as presented in Figure \ref{fig:Microfluidics}. First, a single cylinder of length $l=20$[mm] and diameter of $0.5$[mm] was embedded in an elastic material with dimensions of $10$[mm]x$10$[mm]x$40$[mm] (see Figure \ref{fig:Microfluidics}(1.a)  and additional details in SI). In order to characterize a single viscous peeling based microfluidic channel, we measured flow rate and recorded gap thickness while controlling inlet pressure. The experimental results were compared to steady-state solution of the dimensional version of equation \ref{PME4}, yielding  $q= a^3p_0^4 /192 r_i^2 \pi^2 \mu l $ and $h^4(z)=(1-{z}/{l}){12\mu l a q}/{\pi^2 r_i^2}$, where $p(z=0)=p_0$ and $p(z=l)=0$ are gauge pressures at the inlet and outlet, respectively. Figure \ref{fig:Microfluidics}(1.b) presents the steady state gap at the center of the channel, created by introducing fluid into the layer between the elastic material and solid inner cylindrical core. Figure \ref{fig:Microfluidics}(1.c-1.d) present the experimental measurements compared with the analytic solutions for the flow rate and gap thickness (where $a=2.11[mm^2/MPa]$).  A micron-scale gap is clearly presented and the experimental data agrees well with the simplified model. (We note that in the first actuation after casting, a minimal inlet pressure of about $10$[kPa] is required in order to separate the solids, and the presented results omit this initial actuation.)

The second configuration is a microfluidic network containing 120 parallel channels. Similarly to the single channel configuration, flow rate was measured in order to estimate the gap thickness within the peeling-based channels. The microfluidic device consists of two hollow inlet and outlet outer tubes which are connected with the 120 parallel inner peeling-based channels, as presented in Figure \ref{fig:Microfluidics}(2.a). Inner peeling-based channels with diameter of $\phi_i=0.5$[mm] were fabricated, as well as hollow outer channels with diameter of $\phi_o=2$[mm] (details regarding the fabrication process are presented in the SI section). Since the hollow outer channels have a significantly smaller viscous resistance, pressure variations within the outer channels are neglected in the analytic calculation. Figure \ref{fig:Microfluidics}(2.b) compares the experimental data with the analytic solution and Figure \ref{fig:Microfluidics}(2.c) presents the estimated gap thickness. These results demonstrate the use of viscous peeling to create complex micron-scale network of fluidic channels which were fabricated using low-cost millimeter-scale structures.

The third examined configuration is of an onboard microfluidic valve. Onboard valves are a rather complex yet essential element in microfluidics and lab-on-a-chip devices \cite{unger2000monolithic,thorsen2002microfluidic,desai2012design}. A channel and a valve were fabricated in order to demonstrate the capabilities of viscous peeling based configurations and their compatibility to be integrated in lab-on-a-chip devices. Solid cylindrical core and solid tube-ring shaped valve were embedded in elastomer, as presented in Figure \ref{fig:Microfluidics}(3.a) Fluid at constant pressure of $30$[kPa] was introduced to the channel inlet, yielding a steady state flow in the gap between the solid cylindrical core and the elastic solid. When the valve tube-ring is pressurized, the elastic solid is compressed against the channel's cylindrical core and thus regulates the flow rate, as illustrated in Figure \ref{fig:Microfluidics}(3.b). Valve inlet pressure vs. volumetric flux is presented in Figure \ref{fig:Microfluidics}(3.c), showing the valve's ability to regulate the flow field. This principle can be readily extended to control vast channel networks, with multiple distributed valves. Furthermore, since the propagation of the solid deformation is governed by compactly supported diffusion  equation \ref{eq:final}, multiple valves may be connected to a single inlet, and the actuation of the valves can be sequenced by the propagation of the peeling front between the interconnected valves (see Figure \ref{fig:ModelFull}(d)).

In the following section, the transient effects of the front propagation are examined, in the context of time-dependent deformation of soft actuators.

\section*{Viscous peeling in soft actuators - transient effects}

The emerging field of soft-robotics commonly employs fluidic-driven soft actuators  \cite{ilievski2011soft,onal2016system,onal2017soft}. These soft actuators contain a complex network of interconnected embedded fluid-filled cavities. This section aims to demonstrate fabrication and activation of a viscous peeling based soft actuator, as well as examining transient viscous peeling effects on the time-dependent deformation field of such actuators.  

\begin{figure*}
    \centering

    \includegraphics[scale=0.78]{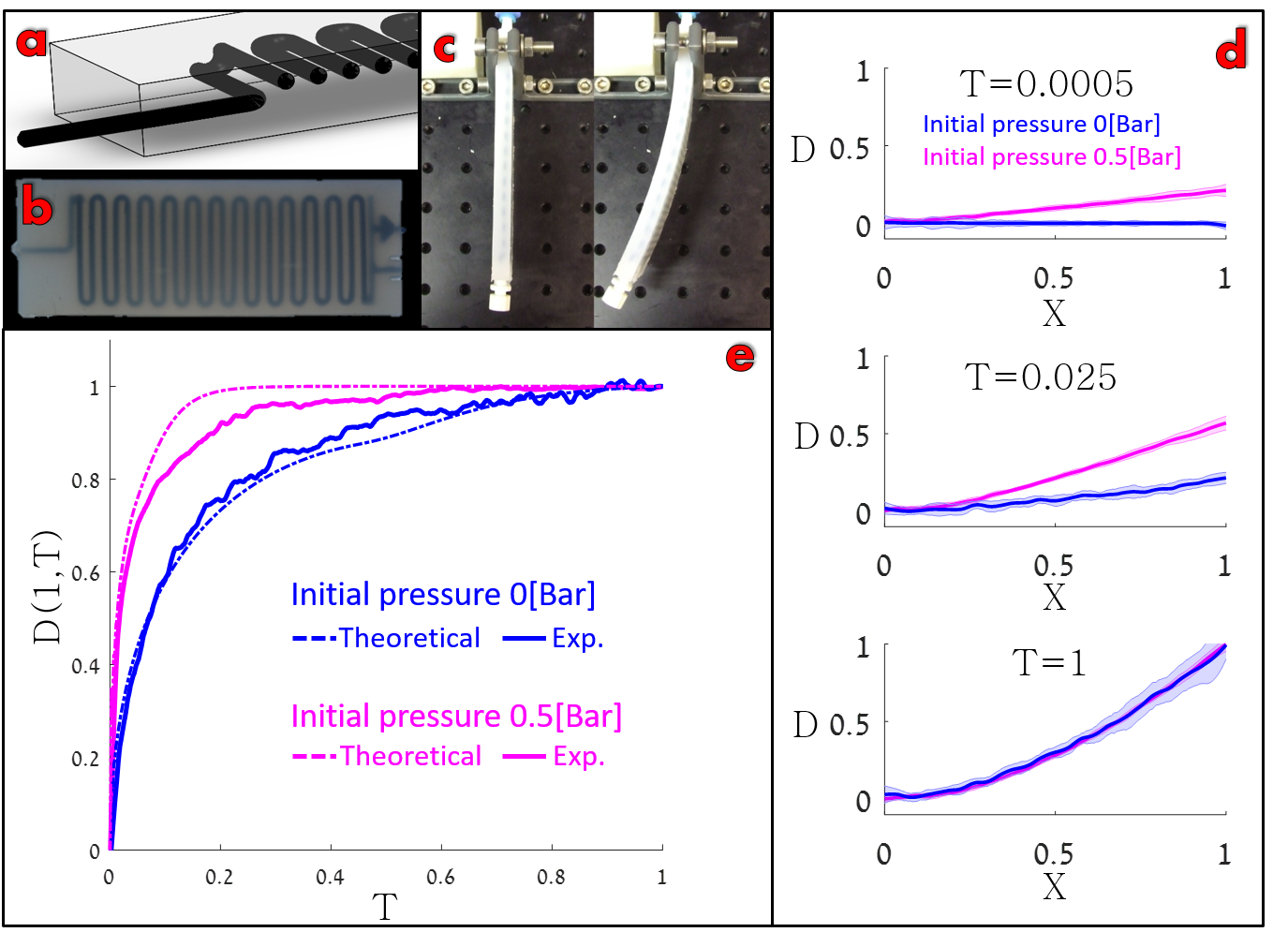}
    \caption{Experimental illustration of a viscous peeling based soft actuator dynamics. \textbf{(a)} CAD image illustrating the cross section of the soft actuator. \textbf{(b)} Top view photo of the actuator showing the embedded serpentine channel. \textbf{(c)} Steady-state unpressurized (left) and pressurized (right) soft actuator. \textbf{(d)} Experimental results of transient actuator's deflection for step response of $0.5$[Bar]. We present two cases: Blue line denotes the deformation of an actuator without initial pressure $P_0=0$[Bar]. Magenta line denotes the deformation of an actuator with initial pressure $P_0=0.5$[Bar] which creates initial fluid layer and increases the channel's cross section area. Shaded areas represents $1$ standard deviation of the experimental measurements. \textbf{(e)} Experimental measurements of the displacement of the actuator's edge $Z=1$ vs time for both cases (solid lines) compared to theoretical model (dashed lines).}
   \label{fig:deformation}
\end{figure*}

Figure \ref{fig:deformation}(a-b) illustrate a viscous peeling based soft actuator with an embedded serpentine core. The core, positioned at an offset from the neutral plane of the beam, is not removed from the elastomer cast. The actuator was mounted in clamped-free configuration, with the deflection plane perpendicular to gravity. Figure \ref{fig:deformation}(c) shows the mounted actuator in unpressurized (left) and pressurized (right) states. The fabrication method and experimental setup are described in detail in the SI section.



Transient actuation of viscous peeling based soft actuators  depends on the speed of propagation of the peeling front, which is strongly affected by the initial wetting layer $\eta_0$. The value of $\eta_0$ can be modified by changing the initial pressure of the fluid. Following previous works \cite{Matia2017,Gamus2017}, the calculation of pressure induced deformation of a beam-shaped soft actuator can be obtained from the modified Euler-Bernoulli equation 
\begin{equation}\label{euler-bernoulli2}
        EI\frac{\partial^2 }{\partial x^2}\left(\frac{\partial^2 d}{\partial x^2}+\lambda \phi(x) p(x,t)\right)+c\frac{\partial d}{\partial t}+\rho_s A\frac{\partial^2 d}{\partial t^2}=q,
\end{equation}
where  $E$  is Young's modulus,  $I$ is second moment of area, $d$ is total beam deformation, $\varphi(x)$ is channel density, $\lambda p$ is beam slope change due to a single channel at pressure $p$,  $c$ is damping coefficient, $\rho_S$  is solid density, $A$ is cross-section area, $q$ is distributed external load and $x$ is an actuator-spatial longitudinal coordinate.

Two cases of transient response to a sudden increase of inlet pressure are presented in Figure \ref{fig:deformation}(d-e). Blue lines mark a configuration where the fluid initial pressure is $P_0=0$[Bar] throughout the network and the suddenly applied inlet pressure is $0.5$[Bar]. In this case the core is initially in contact with the surrounding solid. Magenta lines show a configuration where the initial pressure in the network is $P_0=0.5$[Bar] and the inlet pressure is suddenly increased to $1$[Bar]. In this case, the initial steady pressurization creates solid deformation and a fluid film is contained between the solid core and the elastic material. Figure \ref{fig:deformation}(d) depicts the actuator displacements (relative to the initial state) for both configurations at different times. Figure \ref{fig:deformation}(e) illustrates the position of the actuator tip vs. time for both configurations.  The results show a significant delay between the response of the initially unpressurized configuration (blue lines, $P_0=0$[Bar]) and initially pressurized configuration (magenta lines, $P_0=0.5$[Bar]). This difference represents the nonlinear response of such actuators, where the initial state of the actuator significantly influences the speed of fluid front propagation, and thus the reaction speed to inlet pressurization. The experimental results were compared with the theoretical model  calculated from equations \ref{eq:final} and \ref{euler-bernoulli2} and good agreement is observed.

\section*{Concluding remarks}
Leveraging flow to peel one solid from a second surrounding solid simplifies the fabrication of structures with embedded cavities, while creating a complex and highly nonlinear dynamics in response to fluidic pressurization. This response is governed by a nonlinear diffusion equation relating the viscous resistance of flow within the created cavities to the fluidic pressure. This work presented a model for viscous peeling dynamics, and experimentally demonstrated the properties of such configurations in the context of micro-fluidic networks and fluid-driven soft actuators. 

The viscous-elastic interaction characterizing such structures can be utilized to extend the capabilities of lab-on-a-chip and soft actuators. For example, the distinct peeling front can be used to isolate a propagating fluid in a lab-on-a-chip device, or to clearly differentiate between activated and non-activated regions in the context of soft robotic applications. Additionally, fluidic pressure can be used to create micron-scale geometries from millimeter scale structures. While previous research on viscous peeling type mechanics were mainly examined in the context of geophysical phenomena, the application of such dynamics to lab-on-a-chip and soft-robotics applications creates new fluid mechanics questions. These include viscous peeling of nonuniform internal geometries common in fluidic soft actuators (e.g. embedded sphere- or cube-shaped bladders), or viscous peeling with spatially varying properties of the surrounding elastic material.

\section*{Supplementary Information}
\subsection*{Fabrication and experimental setup of micro-fluidic networks and valves}
The single, peeling-based channel presented in Figure \ref{fig:Microfluidics}(1.a-1.d) was fabricated from ABS fibers via a BCN Sigma 3D printer. The printed cylinder was embedded in Sylgard184 (1:25 ratio) and cured in room temperature for one week. Luer connectors were attached with Sil-Poxy - SmoothOn Inc. The pressure controller was connected to water-filled reservoir connected to the elastic device. We measured inlet pressure (Baumer PBMN B22) and flow rate (CORI-FLOW™, Bronkhorst) while recording the gap thickness via microscope (Nikon eclipse Ti, ANDOR Clara, lumencor Mira).

The network of peeling-based microfluidic channels presented in Figure \ref{fig:Microfluidics}(2.a-2.c) was fabricated using dual-extrusion BCN Sigma 3D printer. In a single print job, inner peeling-based channels were made from PLA fibers, and outer were made from PVA fibers. The single printed part was embedded in  Dragon Skin 30, Smooth-On Inc. Next, only PVA outer channels were dissolved using warm water (40$^{\circ}$ C) creating cavities along the outer channels. The experiment setup is similar to the single channel configuration, excluding a different flow rate indicator (Elveflow MFS5). 

Embedded peeling-based valve presented in Figure \ref{fig:Microfluidics}(3.a-3.c) device was fabricated similarly to the single channel case and the experimental setup utilized the Elveflow MFS5 flow rate indicator and the Baumer PBMN B22 pressure sensor.

\subsection*{Fabrication and experimental setup of viscous peeling based soft actuator}
\begin{figure}
    \centering
    \includegraphics[scale=0.8]{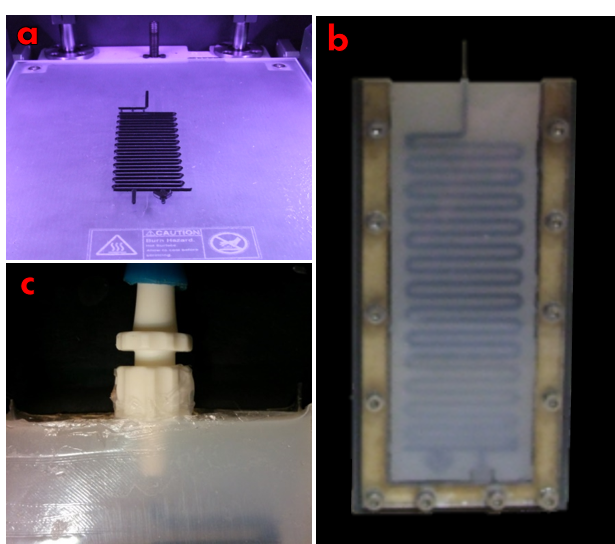}
    \caption{Fabrication process of a viscous peeling based soft actuator. \textbf{(a)} 3D printed PLA core. \textbf{(b)} Elastomer cast around 3D solid core, which remains in the elastomer. \textbf{(c)} Attaching inlet luer connector using Sil-Poxy.}
    \label{fig:Manufacturing}
\end{figure}

\begin{figure}
    \centering
    \includegraphics[scale=0.36]{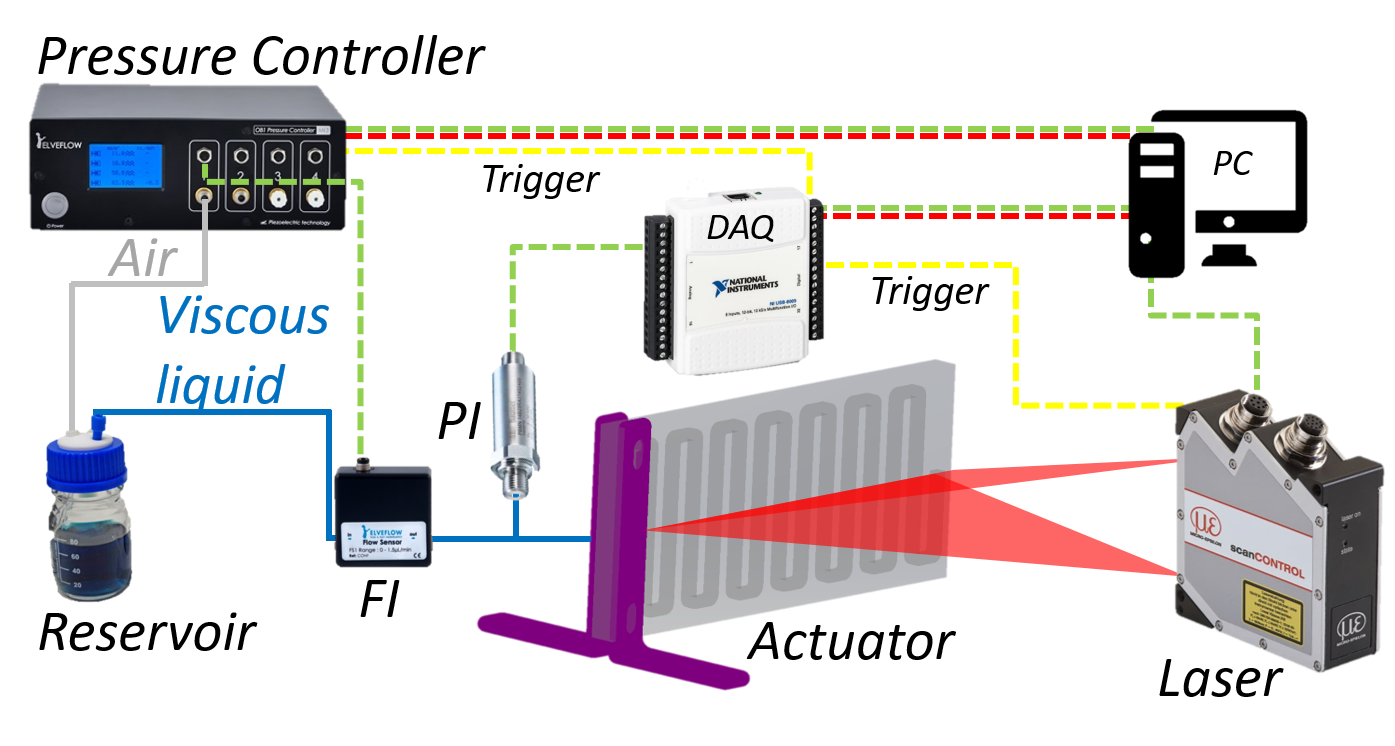}

    \caption{Experimental setup for transient deformation measurements of a viscous peeling based soft actuator. A pressure controller (Elveflow OB1) is connected to fluid-filled reservoir and sets the inlet pressure. Laser profilometer (MicroEpsilon LLT 2650-100) captures and records the time-dependent deformation of the actuator.  Flow rate indicator (Elveflow MFS5) and pressure sensor (Baumer PBMN B22 via NI USB-6009 DAQ) record the flow rate and inlet pressure respectively. Solid lines represent fluidic tube. Dashed lines represent data lines, where red is input, green is output and yellow is trigger.}
    \label{fig:exp_sys}
\end{figure}

The soft actuator presented in Figure \ref{fig:deformation}(3.a-3.c) was fabricated by embedding a 3D printed core (PLA, BCN Sigma) in a serpentine geometry into an elastomer cast (DragonSkin30, Smooth-On Inc.) as shown in Figure \ref{fig:Manufacturing}. The embedded serpentine core is positioned at an offset from the neutral plane of the beam. Luer connectors were glued using silicone rubber adhesive (Sil-Poxy, Smooth-On Inc.). Actuator's geometry and physical properties are: length $l_s=125$[mm], height $h_s=9$[mm], width $b_s=45$[mm], Young's Modulus $E=0.97$ [MPa], density $\rho =1120$[$Kg/m^3$] and damping ratio $\zeta =1.46$. Inner core properties are: length $l_c=1350$[mm], radius $r_c=1$[mm], offset of $2.5[mm]$ from the neutral plane, and the serpentine channel is composed of  $n=26$ parallel channels. The actuator was mounted in clamped-free configuration, with the deflection plane perpendicular to gravity. A pressure controller (Elveflow OB1) was connected to reservoir filled with 60:40 weight ratio glycerin-water  mixture. The reservoir outlet was connected to a flow rate sensor (Elveflow MFS5), pressure sensor (Baumer PBMN B22) and to actuator's inlet. A laser profilometer beam (MicroEpsilon LLT 2650-100) was placed in front of the deflection plane to measure its transverse deformation relative to its contour at rest. The laser was positioned 300[mm] from the actuator and provided 640 spatial sample points. An electric trigger was used to simultaneously activate the pressure controller, flow rate indicator, pressure indicator and the laser profilometer. The experimental setup is illustrated in Figure \ref{fig:exp_sys}.

\subsection*{Experimental relation between fluidic pressure and elastic deformation}

The governing lubrication equations \ref{eq:imass}-\ref{eq:Feta} require an additional relation between the fluidic pressure $P$ and the elastic deformation $\eta$. This relation was estimated experimentally by introducing a pressurized fluid into the gap between a serpentine solid channel and the surrounding elastic material, at pressure range of $0$[kPa] to $100$[kPa]. The geometrical and physical properties of the configuration, as well as the fabrication process, are identical to the previous subsection and described therein. For each fluidic pressure the fluid volume added between the solids was measured visually from a measuring cylinder.

\begin{figure}
    \centering
 
    \includegraphics[scale=0.5]{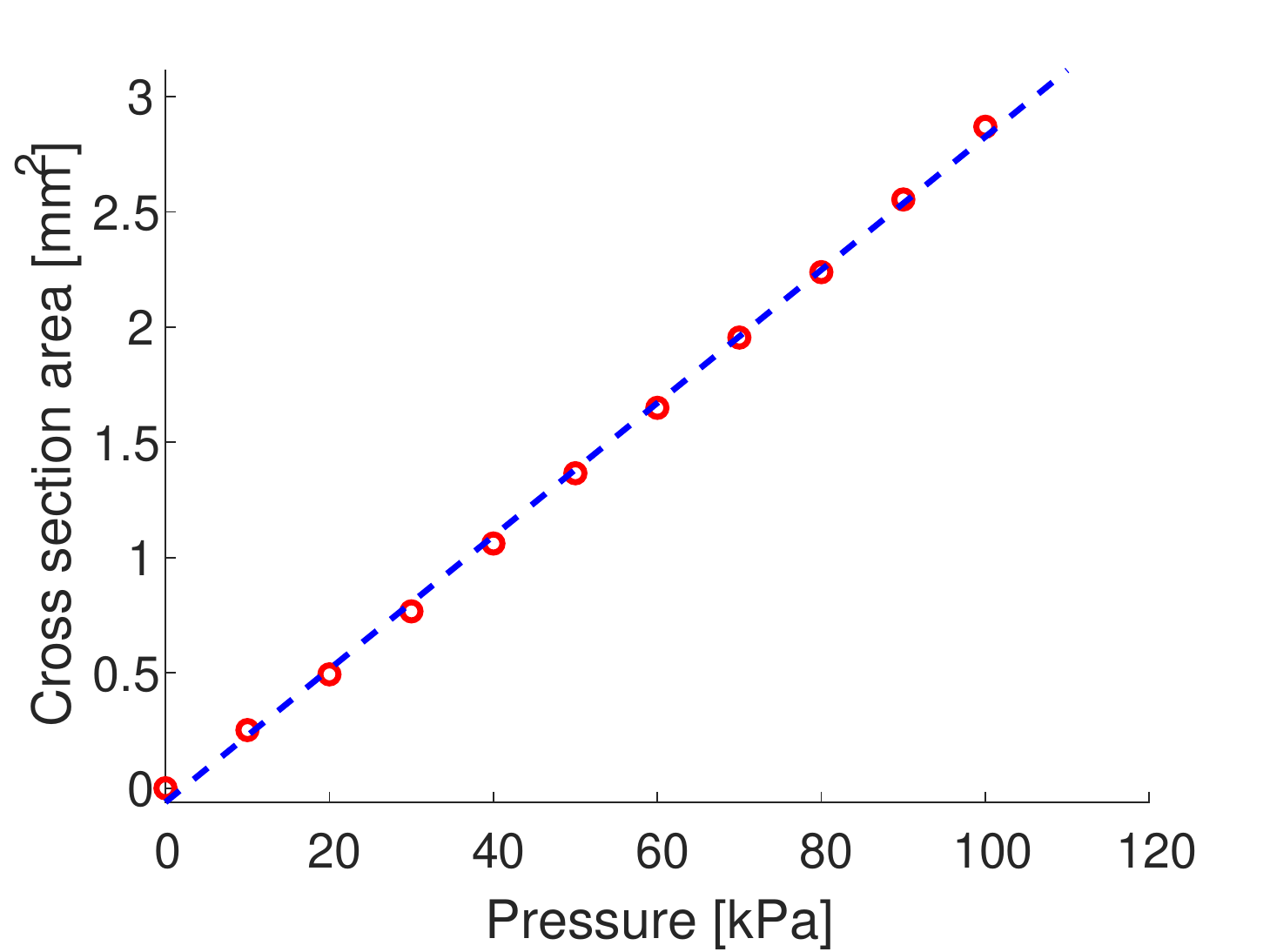}
   
    \caption{Experimental steady-state measurements of fluidic pressure vs variation in cross-sectional area. Cross-sectional area variation is computed by $\Delta v/l$, where $\Delta v$ and the total fluidic volume and $l$ is the total length of the serpentine channel.  Red circles denotes measured values and blue dashed line denotes a linear relation. The geometrical and physical properties of the configuration are described in the text.}
    \label{fig:dadp}
\end{figure}

The variation in cross-sectional area is thus obtained by $\Delta v/l$, where $\Delta v$ and the total fluidic volume and $l$ is the total length of the serpentine channel. Assuming a circular cross-section, the relation between the radial deformation $\eta$ and the fluidic pressure $P$ can thus be obtained. The experimental measurements suggest a linear relation between the pressure $P$, and the change in cross-sectional area, given by 
\begin{equation}
ap=\pi(r_i+h)^2-\pi(r_i+h_0)^2,
\end{equation}
as shown in Figure \ref{fig:dadp}, where $a$ is the coefficient of proportionality, $r_i$ is the inner cylinder radius, $h$ is channel thickness at pressure $p$ and $h_0$ is the channel thickness at $p=0$. In dimensionless form, this relation is given by $AP=(1+\sigma_r\eta)^2-(1+\sigma_r\eta_0)^2$, where $A=a/(\pi r_i^2/E)$. (For sufficiently small $P$, this relation can be simplified further to a linear relation between $P$ and $\eta$.) 

\begin{thebibliography}{9}


\bibitem{grotberg2004biofluid} Grotberg JB, Jensen OE (2004) Biofluid mechanics in flexible tubes. Annual Review of Fluid
Mechanics 36.

\bibitem{michaut2011dynamics} Michaut C (2011) Dynamics of magmatic intrusions in the upper crust: Theory and applications
to laccoliths on earth and the moon. Journal of Geophysical Research: Solid Earth
116(B5).

\bibitem{bunger2011modeling} Bunger A, Cruden A (2011) Modeling the growth of laccoliths and large mafic sills: Role of
magma body forces. Journal of Geophysical Research: Solid Earth 116(B2).

\bibitem{hewitt2015elastic} Hewitt IJ, Balmforth NJ, De Bruyn JR (2015) Elastic-plated gravity currents. European Journal
of Applied Mathematics 26(01):1–31.

\bibitem{thorey2016elastic} Thorey C, Michaut C (2016) Elastic-plated gravity currents with a temperature-dependent
viscosity. Journal of Fluid Mechanics 805:88–117.

\bibitem{makogon2007natural} Makogon Y, Holditch S, Makogon T (2007) Natural gas-hydrates—a potential energy source
for the 21st century. Journal of Petroleum Science and Engineering 56(1-3):14–31.

\bibitem{lai2016elastic} Lai CY, et al. (2016) Elastic relaxation of fluid-driven cracks and the resulting backflow. Physical
review letters 117(26):268001.

\bibitem{young2017long} Young YN, Stone H (2017) Long-wave dynamics of an elastic sheet lubricated by a thin liquid
film on a wetting substrate. Physical Review Fluids 2(6):064001.


\bibitem{hosoi2004peeling} Hosoi AE, Mahadevan L (2004) Peeling, healing, and bursting in a lubricated elastic sheet.
Physical review letters 93(13):137802.

\bibitem{elbaz2016axial} Elbaz SB, Gat AD (2016) Axial creeping flow in the gap between a rigid cylinder and a concentric
elastic tube. Journal of Fluid Mechanics 806:580–602.

\bibitem{vazquez2007porous} Vázquez JL (2007) The porous medium equation: mathematical theory. (Oxford University
Press).

\bibitem{oron1997long} Oron A, Davis SH, Bankoff SG (1997) Long-scale evolution of thin liquid films. Reviews of
modern physics 69(3):931.


\bibitem{huppert1982propagation} Huppert HE (1982) The propagation of two-dimensional and axisymmetric viscous gravity
currents over a rigid horizontal surface. Journal of Fluid Mechanics 121:43–58.

\bibitem{ilievski2011soft} Ilievski F, Mazzeo AD, Shepherd RF, Chen X, Whitesides GM (2011) Soft robotics for
chemists. Angewandte Chemie 123(8):1930–1935.

\bibitem{onal2016system} Onal CD (2016) System-level challenges in pressure-operated soft robotics in SPIE Defense+
Security. (International Society for Optics and Photonics), pp. 983627–983627.

\bibitem{onal2017soft} Onal CD, Chen X, Whitesides GM, Rus D (2017) Soft mobile robots with on-board chemical
pressure generation in Robotics Research. (Springer), pp. 525–540.

\bibitem{Marchese2015} Marchese AD, Katzschmann RK, Rus D (2015) A recipe for soft fluidic elastomer robots. Soft
Robotics 2(1):7–25.

\bibitem{Saggiomo2015} Saggiomo V, Velders AH (2015) Simple 3d printed scaffold-removal method for the fabrication
of intricate microfluidic devices. Advanced science 2(9).

\bibitem{Esser2011} Esser-Kahn AP, et al. (2011) Three-dimensional microvascular fiber-reinforced composites.
Advanced Materials 23(32):3654–3658.

\bibitem{leslie2009frequency} Leslie DC, et al. (2009) Frequency-specific flow control in microfluidic circuits with passive
elastomeric features. Nature Physics 5(3):231.

\bibitem{mosadegh2010integrated} Mosadegh B, et al. (2010) Integrated elastomeric components for autonomous regulation of
sequential and oscillatory flow switching in microfluidic devices. Nature physics 6(6):433.

\bibitem{Leal2007} Leal LG (2007) Advanced transport phenomena: fluid mechanics and convective transport
processes. (Cambridge University Press).

\bibitem{unger2000monolithic} Unger MA, Chou HP, Thorsen T, Scherer A, Quake SR (2000) Monolithic microfabricated
valves and pumps by multilayer soft lithography. Science 288(5463):113–116.

\bibitem{thorsen2002microfluidic} Thorsen T, Maerkl SJ, Quake SR (2002) Microfluidic large-scale integration. Science
298(5593):580–584.



\bibitem{desai2012design} Desai AV, Tice JD, Apblett CA, Kenis PJ (2012) Design considerations for electrostatic microvalves
with applications in poly (dimethylsiloxane)-based microfluidics. Lab on a Chip
12(6):1078–1088.

\bibitem{Matia2017} Matia Y, Elimelech T, Gat AD (2017) Leveraging internal viscous flow to extend the capabilities
of beam-shaped soft robotic actuators. Soft Robotics 4(2):126–134.

\bibitem{Gamus2017} Gamus B, Salem L, Ben-Haim E, Gat AD, Or Y (2017) Interaction between inertia, viscousity
and elasticisy in soft robotic actuator with fluidic network. IEEE Transactions on Robotics
2018, 34, 81.






%
%
%
%
%
%
%

\end{thebibliography}
\end{document}